\def\d{\displaystyle}
\documentclass{article}
\usepackage{graphics,epsf}
\usepackage{amsmath,amssymb}

\begin{document}
\def\thesection{\arabic{section}}
\def\theequation{\arabic{section}.\arabic{equation}}
\hoffset=-20mm \setcounter{page}{1435} SOVIET PHYSICS JOURNAL\hfill Volume 13, Number
11\hfill November 1970\\  \copyright 1973 {\it Consultants Bureau, a division of
Plenum Publishing Corporation, 227 West 17th Street, New York, N.Y. 10011. All rights
reserved. Translated from Izvestiya Vusshikh Uchebnykh Zavedenii Fizika, No. 11, pp.
38-44, November 1970. Original article submitted April 22, 1969; revision submitted
May 11, 1970}.\vspace{5mm}\hrule\vspace{25mm}

\noindent
SHAPE OF THE AMPLIFICATION LINE CORRESPONDING\\
TO AN ADJACENT TRANSITION IN A STRONG FIELD
 \vspace{5mm}

 T. Ya. Popova and A. K. Popov\hfill UDC539.1.01
 \vspace{10mm}

An analysis is made of the effect of a strong field on the shape
of the amplification line for monoenergetic atoms. There are three
strong-field contributions, differing in their dependence on the
set of relaxation parameters of the medium and on the differences
among the populations corresponding to the Raman transitions. An
analysis is made of the conditions under which each of the
contributions is predominant. The change in the line shape by an
intensified external field is found for each case. Neon atoms are
discussed as an example. The results are compared with those
corresponding to a Maxwellian velocity distribution.\\
\\

\section{Introduction}

A strong field which is resonant for one of the allowed
transitions in a medium can change the emission and absorption
spectra corresponding to adjacent transitions [1-3]. This happens
because the photons of a weak field may be emitted in a manner
correlated with that in which photons of the strong field are
emitted [4, 5]. In addition, a strong field changes the
populations and splits the energy levels corresponding to the
resonant transition [6].\\

Each of the three effects has a different dependence on the
unsaturated-level populations and relaxation characteristics of
the medium. These features are displayed most clearly for
uniformly broadened optical transitions, because optical
transitions are characterized by a larger number of relaxation
constants than are microwave transitions, and the population
differences corresponding to the various transitions may differ
markedly. By choosing transitions appropriately one can
suppress some effects while intensifying others.\\

Below we analyze the conditions for the appearance of each of
these effects separately, and we analyze the associated changes in
the spectral properties of the amplification-absorption
coefficient for conditions typical of the optical range.\\

\section{Equation for the Amplification Coefficient}

\setcounter{equation}{0} We consider the amplification (or
absorption) of a weak field at frequency $\omega_{\mu}$,
approximately equal to the frequency $\omega_{gn}$ of the {\it
g-n} transition, in the presence of a strong field $\bf E$ the
frequency of which $\omega$ is approximately equal to the
transition frequency $\omega_{mn}$($E_m > E_g > E_n$). The
weak-field amplification coefficient $\alpha_{\mu}(\Omega_{\mu})$
is related in a simple manner to the emission power per unit
volume $w_{gn}(\Omega_{\mu})$: $$
\alpha_{\mu}(\Omega_{\mu})=\left\{\frac{c}{8\pi}|E_{\mu}|^2\right\}^{-1}w_{gn}
(\Omega_{\mu}),\nonumber $$
where $E_{\mu}$ is the amplitude of the
"weak field", $\Omega_{\mu}= \omega_{\mu}-\omega_{gn}$, and
$w_{gn}$ is calculated as in [3]. We thus have:
\begin{eqnarray}\label{eq1}
\alpha_{\mu}&=&\alpha_{\mu}^{0}\Gamma_{gn}\left\{[\Gamma_{gn}+
i\Omega_{\mu}^{\prime}+|G|^2 (\Gamma_{gm}+i(\Omega_{\mu}^{\prime}-
\Omega^{\prime}))^{-1}]^{-1}\right.\nonumber\\
&&\left.\times\left[1-|G|^2\frac{\Delta n_{mn}/\Delta n_{gn}}{\Gamma^2 (1+\varkappa)+
\Omega^{\prime 2}}\left(\left(1-\frac{\gamma_{mn}}{\Gamma_m}\right)\frac
{2\Gamma}{\Gamma_n}+\frac{\Gamma+i\Omega^{\prime}}{\Gamma_{gm}+i
(\Omega_{\mu}^{\prime}-\Omega^{\prime})}\right)\right]\right\}.
\end{eqnarray}
Here $G= -\, {d_{mn}E}/{2\hbar}$; $\alpha_{\mu}^{0}$ is the
coefficient at the line center in the absence of an external field
($|G|^2 = 0$); $\Gamma_{ik}$, $\Gamma_i$ are the Lorentz
broadenings of the lines and levels ($\Gamma_{mn}\equiv\Gamma$);
$\gamma_m$ is the probability for a transition from level $m$ to
level $n$, $\Omega_{\mu}^{\prime}=\omega_{\mu}-k_{\mu}v$,
$\Omega^{\prime}=\omega-kv$, $v$ is the atomic velocity, $(n_m
-n_n)/(n_g -n_n)\equiv\Delta n_{nm}/\Delta n_{gn}$ is the ratio of
the unsaturated population differences corresponding to the
transitions $m\leftrightarrow n$ and $g\leftrightarrow n$;
$\d\varkappa =(\Gamma_m +\Gamma_n -\gamma_{mn}) \cdot
(\Gamma_m\Gamma_n\Gamma)^{-1}2|G|^2\equiv\tau^2 2|G|^2$. The
population differences $\rho_{mm}-\rho_{nn}$ and $n_g -\rho_{nn}$
depend on the field in the following manner:
\begin{eqnarray}\label{eq2}
\rho_{mm}-\rho_{nn}&=&(\Gamma_2+\Omega^{\prime 2})\Delta n_{mn}[\Gamma^2(1+
\varkappa)+\Omega^{\prime 2}]^{-1},\nonumber\\
n_g -\rho_{nn}&=&\Delta n_{gn}-\Delta n_{mn}\left(1-\frac{\gamma_{mn}}{\Gamma_m}
\right)\frac{2\Gamma}{\Gamma_n}|G|^2[\Gamma^2(1+
\varkappa)+\Omega^{\prime 2}]^{-1}.
\end{eqnarray}
The term proportional to $|G|^2$ in the common denominator in Eq.~(\ref{eq1})
reflects
the broadening and splitting of the line by the strong field; expanding
Eq.~(\ref{eq1}) in simple fractions, we can write the expression for $\alpha_{\mu}$ as:
\begin{eqnarray}\label{eq3}
\alpha_{\mu}&=&\alpha_{\mu}^{0}Re\left\{\frac{\Gamma_{gn}(\alpha_1 -\alpha_2)^{-1}}
{\Gamma_{gn}-\alpha_2^\prime +i(\Omega_\mu^\prime -\alpha_2^{\prime\prime})}
\left(\alpha_1 -|G|^2\frac{\Delta n_{mn}/\Delta n_{gn}}{\Gamma^2(1+
\varkappa)+\Omega^{\prime 2}}\right.\right.\nonumber\\
&&\left.\times\left[\left(1-\frac{\gamma_{mn}}{\Gamma_m}\right)\frac{2\Gamma}
{\Gamma_n}\,\alpha_1-(\Gamma+i\Omega^\prime)\right]\right)-\frac
{\Gamma_{gn}(\alpha_1 -\alpha_2)^{-1}}{\Gamma_{gn}-\alpha_1^\prime
+i(\Omega_\mu^\prime -\alpha_1^{\prime\prime})}\nonumber\\
&&\left.\times\left(\alpha_2-|G|^2\frac{\Delta n_{mn}/\Delta
n_{gn}}{\Gamma^2(1+ \varkappa)+\Omega^{\prime
2}}\left[\left(1-\frac{\gamma_{mn}}{\Gamma_m}\right)
\frac{2\Gamma}{\Gamma_n}\,\alpha_2-(\Gamma+i\Omega^\prime)\right]\right)\right\},
\end{eqnarray}
where
\begin{equation}\label{eq4}
\alpha_{1,2}^\prime+\alpha_{1,2}^{\prime\prime}=
\frac{1}{2}\left(\Gamma_{gn}-\Gamma_{gm}+i\Omega^\prime\pm\sqrt{(\Gamma_{gn}-
\Gamma_{gm}+i\Omega^\prime)^2-4|G|^2}\right).
\end{equation}
It follows from Eqs.(\ref{eq3}) and (\ref{eq4}) that it is easiest
to achieve level splitting with $\Gamma_{gn}=\Gamma_{gm}$, and
this splitting can be observed most easily in its pure form in the
case $n_m =n_n$.\\

The terms proportional to $\d 1-(\gamma_{mn}/\Gamma_{mn})$
describe the changes caused in the populations of levels $m$ and
$n$ by the strong field, and the proportional quantities
$\Gamma+i\Omega^\prime$ corresponds to nonlinear interference
effects [2,3]. The relative weights of these effects depend on
several factors: the relaxation properties of the system, the
ratio $\Delta n_{mn}/\Delta n_{gn}$, and the atomic velocity
distribution. The case of a Maxwell velocity distribution was
analyzed in [3]. Below we take up the case of monoenergetic atoms
(of which a particular case is that of atoms at rest), since all
the effects are displayed most clearly with a uniform broadening
of spectral lines. Below we will omit the primes from
$\Omega^\prime$ and $\Omega_\mu^\prime$ and use $\Omega$ and
$\Omega_\mu$ to signify the deviation from resonance in the
inertial reference system. A real or effective monoenergetic beam
can be produced artificially; an effective beam can be produced,
for example, by exciting atoms with a coherent field from the
ground level to one of the higher-lying levels and by the
subsequent relaxation of the atoms to the $m$, $n$, and $g$ levels.
Conditions can be arranged such that for the levels of interest
the projections of the atomic velocity on the direction of ${\bf
k}_0$ will lie in a very narrow velocity range $\Delta
v=\gamma_0/k_0 \ll\Gamma/k$, $\Gamma_{gn}/k_\mu$ near the velocity
$v_0=\Omega_0/k_0$. Here $\gamma_0$ is the line half-width, and
$k_0$ and $\Omega_0$ are the modulus of the wave vector and the
deviation from resonance for the exciting transition. Then we can
neglect the atomic velocity distributions at the m, n, and g
levels. For a gas with nonuniform broadening these results can be
used to show how the individual atoms interact with the field.
\section{Spectral Properties of the Amplification Coefficient}
\setcounter{equation}{0}
We first take up the case $\Delta n_{mn}=0$, in which the only effect
of the field is to split the levels. Equation (\ref{eq3}) becomes:
\begin{equation}\label{eq3.1}
\frac{\alpha_\mu}{\alpha_\mu^0}=Re\left\{\frac{\Gamma_{gn}}{\alpha_1-\alpha_2}
\left[\frac{\alpha_1}{\Gamma_{gn}-\alpha_2^\prime+i(\Omega_\mu-\alpha_2
^{\prime\prime})}-\frac{\alpha_2}{\Gamma_{gn}-\alpha_1^\prime+
i(\Omega_\mu-\alpha_1^{\prime\prime})}\right]\right\}.
\end{equation}
To determine how the line shape $\alpha_\mu(\Omega_\mu)$ depends on
$|G|^2$ it is sufficient to analyze the following limiting cases.\\

For the case $\Omega =0$ the roots $\alpha_1$ and $\alpha_2$ can be written as:
\begin{eqnarray}\label{eq3.2}
\alpha_1\approx\begin{cases}
(\Gamma_{gn}-\Gamma_{gm})\left(1-\dfrac{|G|^2}{(\Gamma_{gn}-
\Gamma_{gm})^2}\right),&4|G|^2\ll (\Gamma_{gn}-\Gamma_{gm})^2,\nonumber\\
(\Gamma_{gn}-\Gamma_{gm})/2+i|G|,&
4|G|^2\gg (\Gamma_{gn}-\Gamma_{gm})^2;
\end{cases}\\
\alpha_2\approx\begin{cases}
|G|^2/(\Gamma_{gn}-\Gamma_{gm}),&4|G|^2\ll (\Gamma_{gn}-\Gamma_{gm})^2,\\
(\Gamma_{gn}-\Gamma_{gm})/2-i|G|,&4|G|^2\gg (\Gamma_{gn}-\Gamma_{gm})^2.
\end{cases}
\end{eqnarray}
From Eqs. (\ref{eq3.1}) and (\ref{eq3.2}) we see that with $\d
4|G|^2\ll (\Gamma_{gn}-\Gamma_{gm})^2$ the spectrum is a set of
two components having half-widths $\d
\Gamma_{gn}-|G|^2/(\Gamma_{gn}-\Gamma_{gm})$ and $\d
\Gamma_{gm}+|G|^2/(\Gamma_{gn}-\Gamma_{gm})$ The maxima of the two
components occur at the same frequency, but the maximum of the
component having a half-width $\d
\Gamma_{gn}-|G|^2/(\Gamma_{gn}-\Gamma_{gm})$ is higher than that
of the second component by a factor of about $\d
\Gamma_{gm}(\Gamma_{gn}-\Gamma_{gm})^2/\Gamma_{gn}|G|^2$.
Accordingly, the overall effect of a low-intensity external field
in the case $\Omega =0$ is a slight broadening of the spectral
line corresponding to the $g\leftrightarrow n$ transition in the
case $\Gamma_{gm}>\Gamma_{gn}$, or there is a narrowing in the
case $\Gamma_{gm}<\Gamma_{gn}$ For a high-intensity external field
$|G|^2\gg(\Gamma_{gm}-\Gamma_{gn})^2$, the spectrum consists of
two components having the same intensity and the same half-widths
$\d (\Gamma_{gm}+ \Gamma_{gn}/2$. The components are centered at
positions symmetric with respect to the frequency $\Omega_\mu =0$
and are separated by $\Delta\omega =2|G|$.\\

In the other limiting case of $\Gamma_{gm}=\Gamma_{gn}$,
$\Omega\neq 0$ we find:
\begin{eqnarray}\label{eq3.3}
\alpha_1\approx i\Omega(1+|G|^2/\Omega^2),\quad\alpha_2\approx-i|G|^2/\Omega,
\quad 4|G|^2\ll\Omega^2;\nonumber\\
\alpha_1\approx i(\Omega+2|G|)/2,\quad\alpha_2\approx i(\Omega -2|G|)/2,
\quad 4|G|^2\gg\Omega^2.
\end{eqnarray}

\begin{figure}[h]
\center{\epsfxsize=.4\textwidth\leavevmode\epsfbox{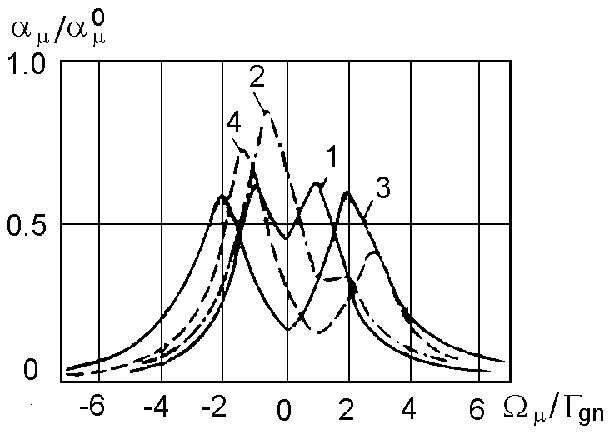}\hspace{12pt}
\epsfxsize=.4\textwidth\leavevmode\epsfbox{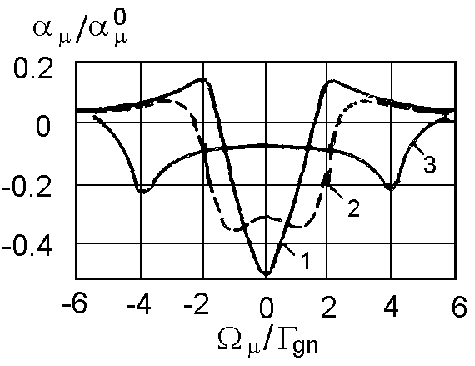}}
\parbox[t]{.47\textwidth}{\caption{
$\Delta n_{mn}=0$.
1) $\varkappa =2,\quad\Omega=0$;
2) $\varkappa=2,\quad \Omega=\Gamma$;
3) $\varkappa=8,\quad\Omega=0$;
4) $\varkappa=8,\quad\Omega=\Gamma.$}}
\hfill
\parbox[t]{.47\textwidth}{\caption{
$x=4.14\quad\Omega=0$. 1) $\varkappa=3,\quad(n_g
-\rho_{nn})/\Delta n_{gn}=0$; 2) $\varkappa=8,\quad (n_g
-\rho_{nn})/\Delta n_{gn}=-0.2$; 3) $\varkappa=40,\quad (n_g
-\rho_{nn})/\Delta n_{gn}=-0.32.$}}
\end{figure}
\begin{figure}[h]
\center{\epsfxsize=.4\textwidth\leavevmode\epsfbox{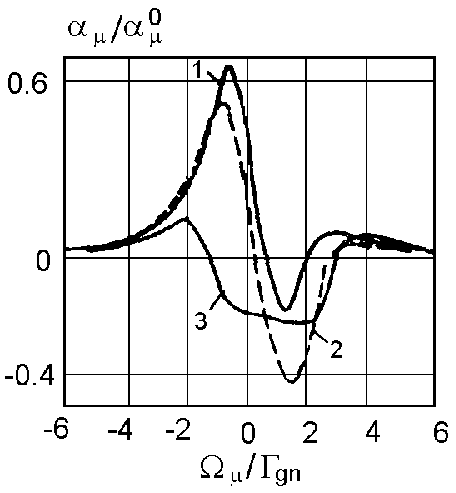}\hspace{12pt}
\epsfxsize=.5\textwidth\leavevmode\epsfbox{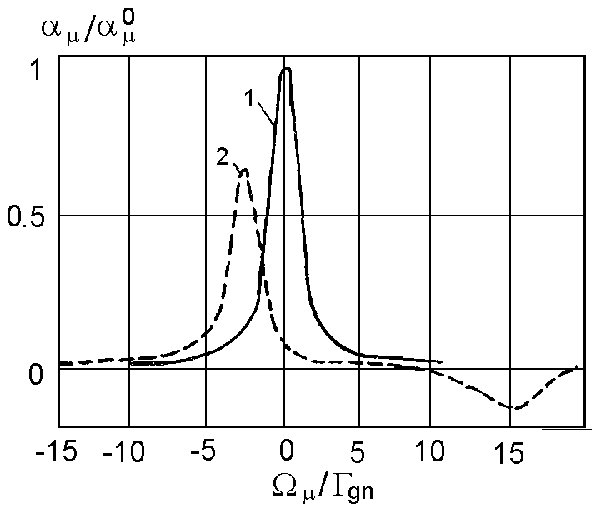}}
\parbox[t]{.47\textwidth}{\caption{
$\Omega=\Gamma,\quad x=4.14$. 1) $\varkappa=1,\quad(n_g
-\rho_{nn})/\Delta n_{gn}=0.55$; 2) $\varkappa=2,\quad(n_g
-\rho_{nn})/\Delta n_{gn}=0.33$; 3) $\varkappa=8,\quad(n_g
-\rho_{nn})/\Delta n_{gn}=-0.08.$ }} \hfill
\parbox[t]{.47\textwidth}{\caption{
$\Omega=10,\quad x=4.14$. 1) $\varkappa=8,\quad(n_g
-\rho_{nn})/\Delta n_{gn}=1$; 2) $\varkappa=100,\quad(n_g
-\rho_{nn})/\Delta n_{gn}=0.33$ }}
\end{figure}

In this case the two spectral components have the same half-width,
$\Gamma_{gn}=\Gamma_{gm}$. For weak fields ($|G|^2\ll\Omega^2$)
one line has a maximum at $\d \Omega_\mu=-|G|^2/\Omega$, while the
other has a maximum at $\d \Omega_\mu=\Omega +|G|^2/\Omega$; the
intensity at the maximum of the first line is higher than that at
the maximum of the center of the second by a factor of $\d
\Omega^2/|G|^2$. In intense fields ($|G|^2\gg\Omega^2$) the two
lines have the same intensity and lie at symmetric positions with
respect to frequency $\d \Omega_\mu=\Omega /2$, separated by
$2|G|$.\\

We can draw the following conclusion regarding the change in the
spectrum accompanying an increase of the intensity of the external
field on the basis of these arguments. When the external field is
applied, we find, in addition to the fundamental component, having
a width of approximately $2\Gamma_{gn}$, an additional component,
having a width approximately equal to that of the line
corresponding to the Raman transition ($2\Gamma_{gm}$), the center
of which is near $\Omega_\mu =\Omega$. As the external field is
intensified, the additional component becomes relatively more
important. The width of each line changes in such a manner that in
the limit of high external field intensities the widths of both
components become the same, equal to $\Gamma_{gn}+\Gamma_{gm}$.
The width change is accompanied by an increase in the separation
between the centers of the spectral components; this separation
depends on both $|G|^2$ and $\Omega$, so that in the limit these
components lie symmetrically about the frequency $\Omega_\mu
=\Omega/2$ separated by $2|G|$. These spectral changes reflect a
modification of the properties of stepped and multiphoton
processes caused by an intensification of the external field [4].
The integral radiation intensity, on the other hand, is governed
only by the change in the quantity $n_g -\rho_{nn}(\varkappa)$
caused by the field and is independent of the external field with
$\Delta n_{mn}=0$. This behavior is illustrated in Fig.1 where the
average values of $\Omega$ and $\varkappa$ for the case of a model
having the relaxation properties of the neon $3s_2 - 2p_4$ and
$2s_2 - 2p_4$ transitions.\\

We turn now to an analysis of the spectral properties of
$\alpha_\mu$, taking into account the interference term
proportional to $\Gamma +i\Omega$. We restrict the discussion to
the case in which the numerator in Eq.~(\ref{eq1}) is governed
primarily by this term, i.e., to the case in which we have $$
\frac{\Gamma}{\Gamma_{gm}}\gg\left(1-\frac{\gamma_{mn}}{\Gamma_m}\right)
\frac{2}{\Gamma_n},\quad
\frac{\Gamma}{\Gamma_{gm}}\cdot\frac{|G|^2\Delta n _{mn}/\Delta
n_{gn}}{\Gamma^2(1+\varkappa)+\Omega^2}\gg 1.$$ {\it In this case
amplification is possible even at $ n_g-\rho_{nn}(\varkappa)<0$
and $\alpha_\mu$ may change sign as a function of
$\bf\Omega_\mu$}. It follows from Eq.~(\ref{eq1}) that with $\d
\Delta n_{mn}/\Delta n_{gn}>0$ amplification occurs in the
frequency band between $(\Omega_\mu)_1$ and $(\Omega_\mu)_2$,
given by
\begin{eqnarray}\label{eq3.4}
(\Omega_\mu)_{1,2}=\dfrac{1}{2\Gamma}\left\{(\Gamma_{gn}+\Gamma
+\Gamma_{gm}) \Omega \pm\sqrt{(\Gamma_{gn}+\Gamma
+\Gamma_{gm})^2\Omega^2-4\Gamma_{gn}\Gamma
\Omega^2+4\Gamma^2(\Gamma_{gn}\Gamma_{gm}+|G|^2)}\right\}.
\end{eqnarray}
It follows from this equation that the band width increases with
increasing $\Omega^2$ and $|G|^2$. In the limiting cases in which
the intense field is far from and close to resonance, we find from
Eq.~(\ref{eq3.4})
\begin{eqnarray}\label{eq3.5}
&(\Omega_\mu)_{1,2}\approx\dfrac{1}{2}\left(1+\dfrac{\Gamma_{gn}}{\Gamma}\right)
\Omega\pm\sqrt{\Gamma_{gn}\Gamma_{gm}+|G|^2},
\hbox{if}\quad\Gamma_{gm}\ll\Gamma,\Gamma_{gn};\quad\Omega^2\ll
\dfrac{4\Gamma^2(\Gamma_{gn}\Gamma_{gm}+|G|^2)}{(\Gamma_{gn}-\Gamma)^2},\nonumber&\\
&(\Omega_\mu)_1\approx\dfrac{\Gamma_{gn}}{\Gamma}\Omega,\quad
(\Omega_\mu)_2\approx\Omega,\quad\hbox{if}\quad\Gamma_{gm}\ll\Gamma,\Gamma_{gn};
\quad\Omega^2\gg
\dfrac{4\Gamma^2(\Gamma_{gn}\Gamma_{gm}+|G|^2)}{(\Gamma_{gn}-\Gamma)^2}.&
\end{eqnarray}
With $\Omega =0$ we have $\d (\Omega_\mu)_{1,2}=\pm
\sqrt{\Gamma_{gn}\Gamma_{gm}+|G|^2}$ for any values of
$\Gamma_{gn}$, $\Gamma_{gm}$ or $|G|^2$. Here the half-width at
half-height of the amplification line is
\begin{eqnarray}\label{eq3.6}
(\Delta\Omega_\mu)_{1,2}^2=&&\dfrac{1}{2}\left\{\sqrt{(\Gamma_{gn}+\Gamma_{gm})^4+
4(\Gamma_{gn}\Gamma_{gm}+|G|^2)^2}-(\Gamma_{gn}+\Gamma_{gm})^2\right\},\nonumber\\
(\Delta\Omega_\mu)_{1,2}^2\approx&&\dfrac{\Gamma_{gn}\Gamma_{gm}+|G|^2}{\Gamma_{gn}+
\Gamma_{gm}},\quad\dfrac{\Gamma_{gn}\Gamma_{gm}+|G|^2}{(\Gamma_{gn}+
\Gamma_{gm})^2}\ll 1.
\end{eqnarray}
It follows from Eqs. (\ref{eq3.5}) and (\ref{eq3.6}) that with
$|G|^2\ll\Gamma_{gn}\Gamma_{gm}$ and $\Gamma_{gm}\ll\Gamma_{gn}$
the width of the amplification band is governed by the geometric
average of $\Gamma_{gn}$ and $\Gamma_{gm}$, while the width of the
amplification line is  $2\Gamma_{gm}$ {\it and may be much narrower
than the natural line width corresponding to the $g\leftrightarrow
n$ transition}. When the frequency of the strong field is scanned,
the amplification band of the weak field also shifts; the band
width depends on both $\Omega^2$ and $|G|^2$.\\

We see from Eq.~(\ref{eq1}) that as field $E$ increases there are
increases in the population and interference contributions to
$\alpha_\mu$. On the other hand, there is a tendency for the
amplification coefficient at the center the line to fall off with
increasing $|G|^2$ because of the level splitting. Analysis of Eq.
(\ref{eq1}) for $\Omega_\mu =\Omega =0$ shows that the optimum
value at $\varkappa$, corresponding to the maximum value of
$\alpha_\mu$ at the line center, is given by $
\varkappa_{opt}=\varkappa_1(x)\{1+\sqrt{1+[x_1\varkappa_1(x)]^{-1}(2\tau^2\Gamma_{gm}
\Gamma_{gn}x+x_1)}\},\quad x>x_1, $ where $ x=\Delta n_{mn}/\Delta
n_{gn}$, $\d x_1 =[(\Gamma_m
-\gamma_{mn}+\Gamma_n\Gamma_m)(2\Gamma_{gm})^{-1}
]^{-1}(\Gamma_m-\gamma_{mn}+\Gamma_n)$, $\d \varkappa_1(x)=(x
x_1^{-1}-1)^{-1}$.\\

The spectral properties of the function $\d \alpha_\mu(\Omega_\mu)/
\alpha_\mu^0$ for the model discussed above are illustrated in
Figs. 2-4, where $x$ is set equal to 4.14, corresponding to the
optimum field $\varkappa =2$. Figure 2 corresponds to the case
$\Omega =0$ for values $\varkappa>\varkappa_{opt}$.  The change in
$\alpha_\mu$ at the maximum  is very rapid while $\varkappa$ varies
near the optimum. (These cases are not illustrated in Fig. 2.) For
example, with $\varkappa=\varkappa_{opt}=2$ we have $\d
\alpha_\mu(0)/\alpha_\mu^0=-32$.  As $\varkappa$ falls off to half
its value, $\d \alpha_\mu(0)/\alpha_\mu^0$ falls off to about
one-third its value. As $\varkappa$ increases to a value 50\%
above $\varkappa_{opt}$, the value of $\d
\alpha_\mu(0)/\alpha_\mu^0$ falls off by a factor of 60. As
$\varkappa$ changes from 2  to 3, the line half-width at
half-height changes from $\approx 0.2\Gamma_{gn}$ to $\approx
\Gamma_{gn}$. With $\varkappa=\varkappa_{opt}$ the $\alpha_\mu$
profile is symmetric and changes sign at
$\Omega_\mu^2\approx\Gamma_{gn}^2$. The absorption in the wings
changes very slowly with increasing $|\Omega_\mu|$. The maximum
absorption is roughly $1/30$ the value
of $|\alpha_\mu(0)|$.\\

Figure 2 shows the change in $\d \alpha_\mu(\Omega_\mu)$
corresponding to a further increase in $\varkappa$. The case
$\varkappa=3$ corresponds to the vanishing of $n_g
-\rho_{nn}(\varkappa)$. The integral value of the coefficient
$\alpha_\mu$ corresponding to $g\leftrightarrow n$ transition also
vanishes. As the external field is intensified, the line splits,
so that with $\varkappa=40$ a plateau appears on the amplification
curve, about $6\Gamma_{gn}$ in width.\\

Figures 3 and 4 show the changes in the spectral properties of the
amplification coefficient as the frequency and intensity of the
strong field are changed. Figure 3 shoves the line profile for
$\Omega=\Gamma$, while figure 4 shows this profile for $\Omega
=10\Gamma$. These cases correspond to quasiresonant Raman
scattering through a common lower level. We see from Figs. 3 and 4
that the frequency separation between the amplification and
absorption maxima increases as the deviation of the strong field
from resonance increases. {\it For fixed $\Delta n_{gn}<0$}, an
increase in $|\Omega|$ requires a more intense external field for
appreciable {\it amplification}. This effect is accompanied by a
change in $n_g -\rho_{nn}(\varkappa)$ which is significant in
comparison with $\Delta n_{gn}$.
\section{Conclusion}
\setcounter{equation}{0} In conclusion we will compare the results
for the cases of a beam having a Maxwell atomic velocity
distribution and a monoenergetic beam. With a Maxwell velocity
distribution, peaks or troughs appear against the background of the
$\alpha_\mu(\Omega_\mu)$ Doppler profile under the influence of
the traveling wave of the intense field; these peaks and dips are
described by
\begin{eqnarray}\label{eq4.1}
&&\dfrac{k_\mu}{k}\dfrac{1\pm\sqrt{1+\varkappa}}{\sqrt{1+\varkappa}}Re\left\{(N_m
-N_n)|G|^2
\left[\Gamma_0+i\left(\Omega_\mu\mp\dfrac{k_\mu}{k}\Omega\right)+
\right.\right.\nonumber\\
&&\left.\left. +\dfrac{|G|^2}{\Gamma_\pm+i\left(\Omega_\mu\mp
\d\dfrac{k_\mu}{k}\Omega\right)}\right]^{-1}\left[\left(1-\dfrac{\gamma_{mn}}
{\Gamma_m}\right)\dfrac{2}{\tilde\Gamma_n}+\dfrac{1}{\Gamma_\pm+i\left(\Omega_\mu\mp
\d\dfrac{k_\mu}{k}\Omega\right)}\right]\right\},
\end{eqnarray}
where
\begin{eqnarray}\label{eq4.2}
k_\mu>k,\quad\Gamma_0=\Gamma_{gn}+\d\dfrac{k_\mu}{k}\Gamma\sqrt{1+\varkappa},\quad
\Gamma_\pm=\Gamma_{gm}+\left(1\mp\dfrac{k_\mu}{k}\right)\Gamma\sqrt{1+\varkappa},\quad
\tilde\Gamma_n=\Gamma_n(1\pm\sqrt{1+\varkappa}).
\end{eqnarray}
The upper sign corresponds to the case $k_\mu k>0$, and the lower
sign corresponds to the case $k_\mu k<0$\,[3]. Comparing Eq.
(\ref{eq4.1}) with Eq. (\ref{eq1}), we conclude that the line
profile for the peaks (or dips) for a gas with a Doppler velocity
distribution is the same as the line shape for an effective
monoenergetic (atomic) beam, for which we have $|N_m -N_n|\gg |N_g
-N_n|,\quad\Omega^\prime =0,\quad\Gamma_{gn}=\Gamma_0,
\quad\Gamma_n =\tilde\Gamma_n,\quad\Gamma_{gm}=\Gamma_\pm$ and for
which the resonant frequency is $\d\omega_{gn}\pm (k_\mu
/k)\cdot\Omega$. However, in the case of a Maxwell distribution the
line shape has a completely different dependence on the magnitude
of the external, field, and there is no profile-asymmetry effect,
as may
be displayed in the case of a monoenergetic beam.\\

The authors thank S. G. Rautian for useful discussions and S. I.
Mortseva for assistance in the numerical calculations.\\
\\
Novosibirk State University. Institute of Semiconductor Physics,
Siberian Branch, Academy of Sciences of the USSR.\\
\\

\section*{LITERATURE CITED}
\begin{enumerate}
\item V. M. Fain, Ya. I. Khanin, and E. G. Yashchin, lzv. VUZ. Radiofiz., {\bf
5}, 697 (1962);
V. M. Fain and Ya. I. Khanin, Quantum Radio Physics [in Russian],
lzd. Sov. Radio (1965).
\item G. E. Notkin, S. G. Rautian, and A. A. Feoktistov, Zh. Eksp. Teor. Fiz.,
{\bf 52}, 1673 (1967).
\item T. Ya. Popova, A. K. Popov, S. G. Rautian, and R. I. Sokolovskii,
Zh. Eksp. Teor. Fiz., {\bf 57}, 85 (1969).
\item H.K. Holt, Phys. Rev. Lett D {\bf 19}, 1275 (1967); ibid {\bf 20}, 410(1968).
\item T. Ya. Popova, A. K. Popov, S. G. Rautian, and A. A. Feoktistov,
Zh. Eksp. Teor. Fiz., {\bf 57}, 444 (1969).
\item  A. K. Popov, Zh. Fksp. Teor. Fiz., {\bf 58}, 1623 (1970).
\end{enumerate}

\end{document}